\documentclass[10pt, conference]{IEEEtran}
\IEEEoverridecommandlockouts
\usepackage{amsmath,amssymb,amsfonts}
\usepackage{algorithmic}
\usepackage[table]{xcolor}
\usepackage{graphicx}
\usepackage{dblfloatfix}
\usepackage{textcomp}

\usepackage{amsmath}
\usepackage{empheq}
\usepackage{float}
\usepackage{multirow}
\usepackage{footnote}
\usepackage{mathtools}
\usepackage{subcaption}
\usepackage{mathtools}
\usepackage[font=small,skip=4pt,belowskip=0pt]{caption}

\usepackage[textsize=tiny,colorinlistoftodos]{todonotes}
\makeatletter
\define@key{todonotes}{bh}[]{
	\setkeys{todonotes}{author=\textbf{Bin - notes}, color=red!30}}%
\define@key{todonotes}{bh2}[]{
		\setkeys{todonotes}{author=\textbf{Bin - urges}, color=green!30}}%
\define@key{todonotes}{zf}[]{
	\setkeys{todonotes}{author=\textbf{Zexin}, color=blue!30}}%
\makeatother

\definecolor{LightGray}{gray}{0.9}

\usepackage{booktabs}
\usepackage{amsthm}

\usepackage[nolist]{acronym}

\usepackage[
	pdfborder={0 0 0},
	]{hyperref}%
	
\begin{document} 

\title{Adaptive Decentralized Queue Disclosure for Impatient Tenants in Edge and Non-terrestrial Systems } 

\author{
	\IEEEauthorblockN{Anthony~Kiggundu\IEEEauthorrefmark{1}\IEEEauthorrefmark{2},~Bin~Han\IEEEauthorrefmark{2}\thanks{B. Han is the corresponding author}~and~Hans~D.~Schotten\IEEEauthorrefmark{1}\IEEEauthorrefmark{2}}
	\IEEEauthorblockA{
		\IEEEauthorrefmark{1}German Research Center for Artificial Intelligence (DFKI), Germany\\
		\IEEEauthorrefmark{2}RPTU University of Kaiserslautern-Landau, Germany\\
	}
}

\maketitle

\begin{abstract}
We study how queue-state information disclosures affect impatient tenants in multi-tenant edge systems. We propose an information-bulletin strategy in which each queue periodically broadcasts two Markov models. One is a model of steady-state service-rate behavior and the other a model of the queue length inter-change times. Tenants autonomously decide to renege or jockey based on this information. The queues observe tenant responses and adapt service rates via a learned, rule-based predictive policy designed for decentralized, partially-observed, and time-varying environments. We compare this decentralized, information-driven policy to the classical, centralized \ac{MDP} hedging-point policy for $M/M/2$ systems. Numerical experiments quantify the tradeoffs in average delay, impatience and robustness to stale information. Results show that when full, instantaneous state information and stationarity hold, the hedging-point policy yields less impatience but this diminishes as information becomes partial or stale. The rule-based predictive policy on the other hand is more robust to staleness in dispatched information, making it conducive for conditions typical of edge cloud and non-terrestrial deployments.
\end{abstract} 


\section{Introduction} 
\label{sec:introduction}
In heterogeneous \ac{5G}/\ac{6G} deployments, multi-tenant resource sharing will be driven not only by static provisioning but increasingly by tenants' autonomous decisions (e.g., whether to offload a task to a remote queue or process it locally). In queuing systems, disclosures of queue state (such as queue lengths, waiting-time estimates or service statistics) can materially alter tenants' behavior and induce competition for resources via jockeying and reneging~\cite{Kiggundu2024ChroniclesOJ,ouyang2022signaling,kiggundu2024}. Queueing control in multi-server systems can be cast either as a centralized optimization problem or as a decentralized behavioral problem where tenants make local decisions using locally available information. \cite{LinBing} generalized the centralized optimal joint routing, service and jockeying policy to have a hedging-point structure. In contrast, modern \ac{MEC} and \ac{NTN} settings are decentralized, suffer partial and stale status broadcasts, and exhibit time-varying channels and mobility. In such environments it is unrealistic to assume a single central controller with instantaneous, global state. 
However, existing disclosure rules and heuristics often developed for static or lightly mobile settings do not, answer three fundamental questions that are central to decentralized control: \textit{what} state information should be shared, \textit{how much} it should be represented, and \textit{how frequently} updates should be dispatched \cite{Kaul2012,Tuhinangshu,Hassin01011994}. These questions are especially salient in self-organizing 6G settings where device heterogeneity, strict \ac{QoS} requirements, and multi-slice subscriptions complicate tenants' preferences and render one-size-fits-all disclosure policies ineffective~\cite{3gpp_slice_orchestra,Gardner}.

We propose and evaluate an \textit{information bulletin} framework in which queues periodically broadcast two Markov models: (i) a service-rate steady-state model (captures processing capability statistics) and (ii) an inter-change time model (captures dynamics of queue length changes)~\cite{mahabhashyam2005queues}. Upon receiving a bulletin (possibly stale, depending on dispatch interval), a tenant decides either to remain, jockey to another queue, or renege and process locally; these micro decisions create observable feedback that queues use to update and learn service-rate policies \cite{Feridun}. Our objective is to minimize a composite measure of system performance that penalizes mean delay, jockeying events and reneging (also referred to as tenant impatience). We show that the resulting optimization formulation is nonconvex and analytically intractable.

Rather than relying on ad-hoc heuristics \cite{ouyang2022signaling} for to optimize the system, we design a rule-based \textit{predictive} policy that (a) estimates service-rate vectors from the two broadcast models, (b) prescribes when and what to broadcast, and (c) adaptively reconfigures queue service rates to regulate impatience. We validate the approach through extensive numerical experiments and sensitivity studies. Our main contributions are:
\begin{enumerate}
  \item We introduce the \textit{information bulletin} concept for multi-tenant queues and formalize two Markov descriptors (service-rate distribution and inter-change time) as dispatchable state summaries suited
for resource-constrained control channels.  
  \item We derive closed-form expressions for jockeying and reneging probabilities under these descriptors, and formulate the joint impatience-minimization problem that optimally trades off delay, jockeying and reneging. We show this optimization is analytically intractable.
  \item We propose a practical, rule-based predictive policy that learns service-rate vectors from tenant responses and adapts the service rates online. 
  \item Our extensive numerical evaluations quantify the value of different bulletin models and dispatch intervals. We also demonstrate robustness of the learned policy across heterogeneous workloads and compare the decentralized, information-driven predictive policy with the hedging-point MDP policy \cite{LinBing}. 
\end{enumerate}

In Section \ref{sec:System Modeling}, we characterize for the information models and the impatience arising from the dispatched information models. We also formulate for a service rate optimization problem and describe our empirical experiments. We then document the numerical evaluations of the information models on the  impatience in Section \ref{sec:results}. We conclude with some open questions and some discussions around possible future work in Section \ref{sec:conlusion}.

\section{System Modeling} 
\label{sec:System Modeling} %
We assumed an $M/M/2$ setup, that is queues $i$ and $j$. Either queue always processes jobs faster than the other. New arrival follow a Poisson distribution with rates $\lambda=\lambda_{i}+\lambda_{j}$. 
The expressions $\mu_{i} = \frac{\lambda_{i} + \delta \lambda} {2}, \mu_{j} = \frac{\lambda_{j} - \delta \lambda} {2}$ (where $0.1<\delta \lambda <0.9$ is a random number) ensure that the rates $\mu_{i}$ and $\mu_{j}$ at which these requests get served vary following a \ac{FCFS} service discipline. 
Each queue dispatches information about its status to tenants at intervals $r$ in seconds introducing some staleness. We characterize for the effect arising from this information staleness using a drift sensitivity $\eta$. 

\subsection{Markov Models of the Queue Status Information }
Our behavioral model is built around two types of information, the Markovian model of the service rates of the queues and the Markovian model of the how the queues change in size, i.e. \ac{ICD}. A round-robin selection technique decides which information model to dispatch during an iteration. 
 
\subsubsection{Markov Model of the Service rates}
Fundamentally, setting high service rates for either queues $i$ or $j$ should lead to the queue shrinking in size faster than when the rates are low which leads to longer queues. The heterogeneity in the service rates also introduces state dependent drifts to yield a scenario of fast and slow service. These interstate visits can result in some bursts in the queue sizes over time.
In equilibrium however, the size of any of these $M/M/1$ queues $i$ or $j$ follows a birth-death \ac{CTMC} with $K$ states and rates $\{\mu_i\}_{i=1}^K$, $\{\mu_j\}_{j=1}^K$ respectively. We let $X$ and $Y$ denote these two stationary service rate distributions over $\{\mu_1<\cdots<\mu_K\}$ with steady state probabilities $\pi_{i}^{X} = Pr\{ \mu = \mu_{i}\}$ and $\pi_{j}^{Y} = Pr\{ \mu = \mu_{j}\}$. Then the effective service rates as averaged over the entire Markov chain are defined by \eqref{eqn:serv_rate}.
\begin{equation}
    \small
    \bar\mu_X = \sum_{i=1}^K \pi_i^X\,\mu_i,
     \quad
    \bar\mu_Y = \sum_{j=1}^K \pi_j^Y\,\mu_j.
    \label{eqn:serv_rate}
\end{equation}

\subsubsection{Markov Model of the Queue Length Dynamics - Inter‐Changing Time Distribution (ICD)}
This measure basically quantifies how frequently transitions occur in a queuing system and can be used to analyze the stability or efficiency of a queuing system. If the queue lengths changes often, then intuitively that queue should be the better one. Fewer perturbations in the system on other hand suggest stability with measurably equal departures and admissions or lower throughput. But low interchange time in the sizes of the queues can also imply underlying performance bottlenecks. 
For the queue therefore in state $n$, when $n=0$ (empty) only an arrival changes the state while when $n\geq 1$ an arrival or departure event can occur to end the queue in state $\lambda_{i}+\mu_{i}$. The Markovian model that encapsulates how frequently the queue length changes in steady state given these events is defined by \eqref{eqn:r_change}.
\begin{equation}
    \small
   R_{i} =\sum_{n=0}^{\infty} \pi_{i,n}\,(\lambda_{i} + \mu_i \cdot 1_{n\geq 1}). 
   \label{eqn:r_change} 
\end{equation}\textit{where $1_{n\geq 1}$ is an indicator function for measurability around $n\geq 1$ since a departure event only occurs when $n\geq 1$}.
However, \eqref{eqn:r_change} can be simplified as
\begin{equation}
    \small
     \sum_{n=0}^{\infty} \pi_{i,n}\,(\lambda_{i} + \mu_i \cdot 1_{n\geq 1}) = \pi_{i,0}\lambda_{i} + \sum_{n=1}^{\infty} \pi_{i,n}\,(\lambda_{i} + \mu_{i}),
    \label{eqn:expanded}
\end{equation} 
where $\pi_{i,0} = 1 -\rho_{i}$ and $\sum_{n=1}^{\infty} \pi_{i,n} = \rho_{i}$ and eventually:
\begin{equation}
    \small
       R_{i} = (1-\rho_{i})\lambda_{i} + \rho_{i}(\lambda_{i}+\mu_{i})
        = \lambda_{i}+ \rho_{i}\mu_{i}.   
\end{equation}
Since $\rho_{i}\mu_{i} = \lambda_{i}$,
\begin{equation}
    \small
       R_{i} = \lambda_{i} + \lambda_{i} = 2\lambda_{i}
       \label{eqn:swings}
\end{equation} is a count of the number of changes introduced by these events.  We can therefore express for the expected time between successive changes in the sizes of the queue $i$ or $j$ using \eqref{eqn:change_time}.
\begin{equation}
    \small
    T^{i}_{\text{ICD}}= \frac{1}{R_{i}} = \frac{1}{2\lambda_{i}}
    \label{eqn:change_time}
\end{equation}
\subsubsection{First-Order Stochastic Dominance (FSD)}
To find the better queue given its service rate distribution, we need to characterize for the cumulative density functions (CDF) $F_{X}(\mu_{k})$ and $F_{Y}(\mu_{k}) $ ($k=1,\ldots , K$) of $X$ and $Y$ from the steady state probabilities $\pi_{i}^{X}$, $\pi_{j}^{Y}$ above using \ref{eqn:redefcdf}. 
\begin{equation}
   \small
    F_{X}(\mu_{k}) =  \sum_{i=1}^{K} \pi_{i}^{X}, \quad F_{Y}(\mu_{k}) = \sum_{j=1}^{K} \pi_{j}^{Y}
    \label{eqn:redefcdf}
\end{equation}
Then, according to \cite{torres1990stochastic}, if $P[X>x] \geq P[Y>x]$ $\forall{x\in R}$ then $X$ is said to first order stochastically dominate $Y$ and the conditions defined by \eqref{eqn:fsd} must hold.
\begin{equation}
\small
  \begin{split}
     F_{X}(\mu_{k}) \leq  F_{Y}(\mu_{k}) \quad \forall{k=1,\ldots ,K} \\
     \exists k \quad F_{X}(\mu_{k}) < F_{Y}(\mu_{k})
  \end{split}   
    \label{eqn:fsd}
\end{equation}
Therefore, the rationale to abandon the system or switch queues given the comparison of the the service rates distributions is defined around the stochastic dominance of one distribution against the other as conditioned by \eqref{eqn:fsd_test}.
\begin{equation}
  \small
   \sum_{i=1}^{K} \pi_{i}^{X} \leq \sum_{i=1}^{K} \pi_{i}^{Y}, \quad \forall\,k,
 \quad \exists\,k:\,F_X(\mu_k)<F_Y(\mu_k).
    \label{eqn:fsd_test}
\end{equation} \textit{where $ k=1,\ldots, K$}

\subsection{Reneging Behavior}
The decision to abandon the queue is premised on the condition that the local processing $T_{local}$ (deterministic and no waiting time involved) is less than the estimated waiting time in either queues, such that jockeying to the alternative queue is not an option. However, FSD is a comparison between two distributions of the service rate vectors of queues $i,j$. From \eqref{eqn:serv_rate} and \eqref{eqn:redefcdf} therefore, queue $i$ FSD-dominates queue $j$ when $F_{i}(\bar\mu_{i}) \geq F_{{j}}(\bar\mu_{j}) \quad \forall{t\geq 0}$. Given the better service rate vector, we need to redefine its corresponding CDF in terms of the waiting times distribution $W_{i} \sim f_{w_{i}}(t) $ or $W_{j} \sim f_{w_{j}}(t)$ in steady-state $\pi_{j}(n)$ before direct comparison to $T_{local}$ is made. In equilibrium, for a buffered request in queue $i$ at position $\ell$, we express for the expected remaining time until that request gets served using \eqref{eqn:remaining}.  
\begin{equation}
    \small
    \mathbb{E}[W_{i}|\ell] = \sum^{\ell}_{i=0}\frac{1}{\mu_{i}} = \frac{\ell}{\mu_{i}}
    =\sum_{i=0}^{\infty}\pi_{\ell}\,\frac{\ell}{\mu_{i}}
    \label{eqn:remaining}
\end{equation} And the \ac{CDF} of this remaining time generally obeys the Erlang distribution,$W_{i}| l \sim \text{Erlang}(l,\mu_{i})$ whose definition takes the form of \eqref{eqn:remain_erlang}.
\begin{equation}
    \small
    F_{W_{i}|\ell}(t) = \mathbb{P}(W_i \le t \mid \ell)
= 1 - \sum_{v=0}^{\ell-1} \frac{[(\mu_{i}- \lambda_{i} )t]^v}{v!} e^{-(\mu_{i}- \lambda_{i}  )t}  
  \label{eqn:remain_erlang}
\end{equation}
But the reneging tenant in queue $i$ first weighs the option whether to jockey and land at position $k$ in the other queue $j$. The expected sojourn time of a jockeyed request to queue $j$ is also generalized to an Erlang distribution of order $k$ ($W_{j}\sim \text{Erlang}(k,2\mu_{j})$) and its \ac{CDF} is expressed for by \eqref{eqn:erlang}. 
\begin{equation}
    \small
   F_{W_{j}|k}(t) = \mathbb{P}(W_{j} \le t | k) = 1 - \sum_{v=0}^{k-1} \frac{[(2\mu_j - \lambda_j)t]^v}{v!} e^{-(2\mu_j - \lambda_j)t}
    \label{eqn:erlang}
\end{equation}
Then taking the mean over all $n$ requests under stationary conditions $\pi_{n}$ we redefine the CDFs of the expected waiting times of either queues as:
\begin{equation}
\small
\begin{split}
    F_{W_{i}}(t) = \sum_{\ell=0}^{\infty} \pi(n) \cdot \mathbb{P}(W_i \leq t \mid \ell) \\
    F_{W_{j}}(t) = \sum_{n=0}^{\infty} \pi(n) \cdot \mathbb{P}(W_j \leq t \mid n)
\end{split}   
\end{equation}
Therefore based on the comparison $F_{W_{i}}(t) \geq F_{W_{j}}(t) \quad \forall{t\geq 0}$ and assuming queue $i$ is the better queue at pose $\ell$, \eqref{eqn:compare} characterizes the decision to renege to local processing as a final comparison to the local processing $T_{local}$ otherwise jockey to the better queue.
\begin{equation}
    \small
   \begin{split}
     P^{\text{FSD}}_{reneg}(\ell) = \mathbb{P}(W_{i}>T_{local}|\ell)    
     = 1 -F_{W_{i}|\ell}(T_{local})
   \end{split} 
   \label{eqn:compare}
\end{equation} 

Factoring in the interval $r$ at which the status is dispatched and the staleness factor $\eta$, the reneging probability $P^{\text{FSD}}_{reneg}$ given the FSD of the better of two queues in comparison to local processing is then:
\begin{equation}
\small
\begin{split}
\mathbb{P}^{\mathrm{FSD}}_{\rm reneg}(\ell)
    &= 1 - \Bigl[\,1 -
      \sum_{v=0}^{\ell-1}
      \frac{\bigl[(\mu_{i}-\lambda_{i})\,\Delta\bigr]^{v}}{v!}
      \,e^{-(\mu_{i}-\lambda_{i})\,\Delta}
      \Bigr] \\[4pt]
    &= \sum_{v=0}^{\ell-1}
     \frac{\bigl[(\mu_{i}-\lambda_{i})\,\Delta\bigr]^{v}}{v!}
     \;e^{-(\mu_{i}-\lambda_{i})\,\Delta},
 \end{split}
 \label{eqn:probab_renege}
\end{equation}
\textit{where $\Delta = T_{local}- \eta r$ ,$r$ as the dispatch interval, $\eta\in[0,1]$ denoting how degraded the status is, i.e. perfect state versus staleness.}

\noindent And the rate at which requests that renege from a particular queue when consuming from the FSD information source is calculated using \eqref{eqn:rate_renege}.
\begin{align}
    R^{\mathrm{FSD}}_{\mathrm{reneg}}(\ell)
    &= \lambda_{i}\sum_{\ell=0}^{\infty}\pi_{\ell}\;P^{\mathrm{FSD}}_{\mathrm{reneg}}(\ell) \\[6pt]
    &= \lambda_{i}\sum_{\ell=0}^{\infty}\pi_{\ell}
   \sum_{v=0}^{\ell-1}
     \frac{\bigl[\mu_{i}\,(T_{\mathrm{local}}-\eta\,r)\bigr]^v}{v!}
                   e^{\bigl[-\mu_{i}\,(T_{\mathrm{local}}-\eta\,r)\bigr]}
    \label{eqn:rate_renege}
\end{align}

On the other hand, dispatching the interchanging times of the queue lengths implies the rationale to abandon the current queue $i$ or $j$ for local processing follows from a comparison of how often the sizes of both queues change. The comparison $T^{i}_{\text{ICD}} < T^{j}_{\text{ICD}}$ gives the queue whose length is changing often. This could imply more impatient tenants or more requests being processed. Assuming $T^{i}_{\text{ICD}} < T^{j}_{\text{ICD}}$ is true, then the tenant compares the remaining time at position $\ell$ in the current queue $i$ defined by \eqref{eqn:remaining} to $T_{local}$. Such that, the probability $P^{\mathrm{ICD}}_{\mathrm{reneg}}(\ell)$ to renege to local processing is similarly defined using \eqref{eqn:probab_renege}.
And the percentage $R^{\text{ICD}}_{reneg}$ of requests that renege from the queue given this kind of information is also defined using 

\begin{equation}
 \small
   R^{\text{ICD}}_{reneg} = \lambda_{i}\sum^{\infty}_{n=1}\pi_{n}\cdot P^{\text{ICD}}_{reneg}(n,\mu_{i},(T_{local}-\eta\,r))
    \label{eqn:rene_it_rate}
\end{equation}

\subsection{Jockeying Behavior}
From \eqref{eqn:swings}, it is known that the rate at which the sizes of either queues changes is characterized by the arrivals and departures. We take the case of independence of both events here and denote as $M_{i}, M_{j}$ the number of such events in queue $i,j$ respectively. Then a request switches from queues $i$ to $j$ given the comparison between the interchanging times $T^{i}_{\text{ICD}}$ and $T^{j}_{\text{ICD}}$ as defined by \eqref{eqn:change_time}. The probability $P^{\text{ICD}}_{i \to j}$ therefore to jockey from queue $i \to j$ and vice-versa is defined around this difference in the number of events in either queues as:
\begin{equation}
 \small
   P^{\text{ICD}}_{i \to j} = \sigma(d[M_{i}-M_{j}])    
\end{equation}
\textit{where $\sigma=\frac{1}{1+e^{-x}}$ is a sigmoid function that tunes how steep the probability rises with the drift difference and $d>0$ is a decision function that regulates the switching behavior given this difference.}

From \eqref{eqn:swings}, we can redefine $M_{i}= 2\lambda_{i}e^{-\eta r}$ and $M_{j}=2\lambda_{j}e^{-\eta r}$, then based on the difference $M_{j} - M_{i} = 2(\lambda_{j} - \lambda_{i})e^{-\eta r}$ the jockeying probability is characterized for using \eqref{eqn:violate_fsd}
\begin{equation}
 \footnotesize
    \begin{split}    
      P^{\text{ICD}}_{i\to j} = \sigma(d[2\lambda_{i}e^{-\eta r} - 2\lambda_{j}e^{-\eta r}]) \\        
      \quad  = \frac{1}{1+e^{-2d e^{-\eta r}(\lambda_{i}-\lambda_{j})}}       
      \end{split}
    \label{eqn:violate_fsd}
\end{equation}\textit{since $\lambda_{i} = \mu_{i}\rho_{i}, R_{i}=2\lambda_{i} = 2\mu_{i}\rho_{i}$}.

\noindent The jockeying rate within the system in this case is then defined for by:
\begin{equation}
 \footnotesize
    R^{\text{ICD}}_{i\to j} = \lambda_{i}P^{\text{ICD}}_{i\to j} + \lambda_{j} P^{\text{ICD}}_{j\to i}
\end{equation}
 
In the case of the FSD-based waiting time distribution, \eqref{eqn:remain_erlang} characterizes for the CDF of the remaining waiting time $W_{i}|l$ at position $l$. Similarly, \eqref{eqn:erlang} is definitive of the expected waiting time when this request is migrated to queue $j$. The decision to move from $i \to j$ is therefore pegged on the FSD comparative rule $F_{W_{i}}(t) \geq F_{W_{j}}(t) \quad \forall{t}$. 
\begin{equation}
 \small
    \begin{split}
        \mathbb{P}_{\,i\to j}^{\mathrm{FSD}}(\ell)
        &=
        \mathbb{P}\{W_{j,k}<W_{i,\ell}\}
        =
        \int_{0}^{\infty}
        f_{W_{j,k}}(t)\,
        \bigl[1 - F_{W_{i,\ell}}(t)\bigr]\,dt,
    \end{split} 
    \label{eqn:no_staleness}
\end{equation}\textit{where $f_{W_{j,k}}(t)= \frac{d}{dt}F_{W_{j,k}}(t)$ is the probability density function in $j$ and $[1 - F_{i,\ell}(t)\bigr]$ corresponds to $\mathbb{P}(W_i>t)$}

\noindent But because the states are dispatched every $r$ seconds, the degradation in updates suffers a staleness. And the probability that the request is moved to from queue $i \to j$ is defined using \eqref{eqn:fsd_probab_jockey}
\begin{equation}
 \footnotesize
    \begin{split}
        \mathbb{P}_{\,i\to j}^{\mathrm{FSD}}(\ell)
        &=
        \int_{0}^{\infty}
        \frac{\bigl[\xi_j\,(t-\eta r)\bigr]^{k-1}}{(k-1)!}\,\xi_j\,
        e^{-\xi_j\,(t-\eta r)}\\
        &\qquad\quad\times
        \sum_{m=0}^{\ell-1}
        \frac{\bigl[\xi_i\,(t-\eta r)\bigr]^{m}}{m!}\,
        e^{-\xi_i\,(t-\eta r)}\,dt,
    \end{split}
    \label{eqn:fsd_probab_jockey}
\end{equation}
Here, $\xi_{i} = 2\mu_{i}-\lambda_{i}$ and $ \xi_{j} = 2\mu_{j}-\lambda_{j}$ and \eqref{eqn:fsd_probab_jockey} which can be expressed as:
\begin{equation}
 \footnotesize
    \mathbb{P}_{\,i\to j}^{\rm FSD}(\ell)
      =\sum_{k=0}^{\ell-1}
      \frac{\xi_j^\ell}{(\xi_i+\xi_j)^{\,\ell+k}}
       \;\frac{(\ell+k-1)!}{\,(\ell-1)!\;k!}\,
\end{equation}
And in stable conditions, \eqref{eqn:fsd_jockey_rate} characterizes for the rate a which requests are moved around in time.
\begin{equation}
     R^{\text{FSD}}_{i \to j} = \lambda_{i}P^{\text{FSD}}_{i\to j} + \lambda_{j}P^{\text{FSD}}_{j\to i}
    \label{eqn:fsd_jockey_rate}
\end{equation}
Given these information models, the queues are expected to adapt to the impatient tenant's behavior and learn to recalibrate their service rate vectors $\bar\mu$ to minimize average delay and the overall impatience among the tenants. 
This translates into an optimization problem, where we can jointly choose the service‐rate vector $\bar\mu$ in \eqref{eqn:serv_rate} that satisfies the \ac{KKT} conditions. We limit the scope of the formulation to the $\text{ICD}$ information (although the characterization for the $\text{FSD}$ would a adopt similar form) model and \eqref{eqn:optimize} is definitive of the corresponding optimization problem.
\begin{equation}\label{eqn:optimize}
    \small
    \begin{aligned}
        \min_{\mu_i,\mu_j}\ &\tau\bigl[W_i(\mu_i)+W_j(\mu_j)\bigr]
        +\varphi\bigl[R_i^{\rm reneg}(\mu_i)+R_j^{\rm reneg}(\mu_j)\bigr]\\
        &+\psi\bigl[R_{i\to j}^{\rm jockey}(\mu_i,\mu_j)+R_{j\to i}^{\rm jockey}(\mu_j,\mu_i)\bigr]\\
        \text{s.t.}\ &\mu_{i,\min}\le\mu_i<\mu_{i,\max},\ \mu_i>\lambda_i,\\
        &\mu_{j,\min}\le\mu_j<\mu_{j,\max},\ \mu_j>\lambda_j
    \end{aligned}
\end{equation}
\textit{where $W_{i(\mu_{i})} = \frac{\rho_{i}}{\mu_{i} - \lambda_{i}}$} 

\noindent The corresponding optimization function to this problem is given by \eqref{eqn:obj_function}
\begin{equation}\label{eqn:obj_function}
    \footnotesize
    \begin{aligned}
        f(\mu_i,\mu_j)
        &=\tau\Big(\frac{\rho_i}{\mu_i-\lambda_i}+\frac{\rho_j}{\mu_j-\lambda_j}\Big)\\
        &\quad+\varphi\big(\lambda_iR^{\rm ICD}_{\rm reneg}(\mu_i)+\lambda_jR^{\rm ICD}_{\rm reneg}(\mu_j)\big)\\
        &\quad+\psi\Big(\lambda_i\sigma\big(d[2\lambda_i e^{-\eta r}-2\lambda_j e^{-\eta r}]\big)\\
        &\qquad\qquad+\lambda_j\sigma\big(d e^{-\eta r}[(\mu_i-\lambda_i)-(\mu_j-\lambda_j)]\big)\Big).
    \end{aligned}
\end{equation}

\noindent When there are $n=0$ requests in the system, the independence between arrival and a departure events implies $\epsilon_{i\neq j}= \mu_{i}-\lambda_{i}$ since the two events happening simultaneously ends the queue in an empty state. We take the independence of events to simplify the third term in \eqref{eqn:obj_function} as $z = d e^{-\eta r}[\epsilon_{j} - \epsilon_{i}]$, such that $\sigma'(z)=\sigma(z)\bigl[1-\sigma(z)\bigr]$.
We then define the inequality constraints $g_{1,i}(\mu_{i}) \;=\;\mu_{i,\min} - \mu_{i} \;\le\;0,\quad
g_{2,i}(\mu_{i}) \;=\;\mu_{i} - \mu_{i,\max} \;\le\;0,\quad
g_{3,i}(\mu_{i}) \;=\;\lambda_{i} - \mu_{i} \;\le\;0 $ for both $i$ and $j$.
Equation \eqref{eqn:lagrange} expresses for the corresponding Lagrange Function with multipliers $ \gamma_{m,i}\ge 0 $ for each of $m=3$ constraints $g_{m,i}\leq 0$ and $g_{m,j}\leq 0$.
\begin{equation}\label{eqn:lagrange}
    \footnotesize
    \begin{aligned}
        \mathcal L(\mu_i,\mu_j,\{\gamma_{k,i}\})
        &=f(\mu_i,\mu_j)\\
        &\quad+\sum_{i=1}^2\Big[\gamma_{1,i}(\mu_{i,\min}-\mu_i)
        +\gamma_{2,i}(\mu_i-\mu_{i,\max})\\
        &\qquad\qquad+\gamma_{3,i}(\lambda_i-\mu_i)\Big].
    \end{aligned}
\end{equation}

\noindent In equilibrium,  $\nabla_{\mu_{i},\mu_{j}}\mathcal{L}=0$, which implies that for both $i$ and $j$, we express for the derivatives of each term in \eqref{eqn:obj_function} as:
\begin{equation}
 \footnotesize
   \begin{split}
       \frac{\partial f}{\partial \mu_{i}} - \gamma_{1,i} + \gamma_{2,i} + \gamma_{3,i} = 0 \\ 
       \frac{\partial f}{\partial \mu_{j}}- \gamma_{1,j} + \gamma_{2,j}+ \gamma_{3,j} = 0
   \end{split} 
\end{equation}

Such that \eqref{eqn:partials_ders_1} and \eqref{eqn:partials_ders_2} characterize for the KKT stationarity conditions.
\begin{equation}
 \small
  \begin{aligned}
    0
    &=
    -\tfrac{\lambda_i(2\mu_i-\lambda_i)}{\mu_i^2(\mu_i-\lambda_i)^2}
    \;+\; 
    \varphi\,\lambda_i\sum_{v=0}^{\ell_i-1}\Bigl[\tfrac{v}{\epsilon_i}-\Delta\Bigr] 
    \frac{(\epsilon_i\Delta)^v}{v!}e^{-\epsilon_i\Delta}\\
    &\quad
    +\;\psi\,d\,e^{-\eta r}\,\sigma'(z)\,(\lambda_i+\lambda_j) 
    \;-\;\gamma_{1,i} + \gamma{2,i} + \gamma{3,i} .
  \end{aligned} 
  \label{eqn:partials_ders_1}
\end{equation} and 
\begin{equation}
 \footnotesize
  \begin{aligned}
    0
    &=
    -\tfrac{\lambda_j(2\mu_j-\lambda_j)}{\mu_j^2(\mu_j-\lambda_j)^2}
    +\; 
    \varphi\,\lambda_j\sum_{v=0}^{\ell_j-1}\Bigl[\tfrac{v}{\epsilon_j}-\Delta\Bigr] 
    \frac{(\epsilon_j\Delta)^v}{v!}e^{-\epsilon_j\Delta}\\
    &\quad
    -\;\psi\,d\,e^{-\eta r}\,\sigma'(z)\,(\lambda_i+\lambda_j) 
    \;-\;\gamma_{1,j} + \gamma_{2,j}+ \gamma_{3,j}.
  \end{aligned}
  \label{eqn:partials_ders_2}
\end{equation}
Essentially, from these formulations, we can randomly select an active-set of multipliers, resolve the nonlinear equations \eqref{eqn:partials_ders_1},\eqref{eqn:partials_ders_2} and perform both complementary slackness and feasibility tests for $i,j$ using 
\begin{equation}
\small
\begin{split}
     \gamma_{1,i}\,(\mu_{i,\min}-\mu_{i})=0, \quad  
   \gamma_{2,i}\,(\mu_{i}-\mu_{i,\max})=0,  \\
    \gamma_{3,i}\,(\lambda_{i}-\mu_{i})=0, \quad 
    \gamma_{1,i}\ge0,\;\gamma_{2,i}\ge0,\;\gamma_{3,i}\ge0
\end{split}  
   \label{eqn:slackness}
\end{equation}

Any $(\mu_{i}^{*},\mu_{j}^{*},\{\gamma_{j,i}^{*}\})$ satisfying all of these is a candidate local minimum. If the solution lies outside the bound defined by \eqref{eqn:slackness}, we validate the KKT conditionality under defined constraints after a projection to boundaries. However, $f$ must be jointly convex in $(\mu_{i},\mu_{j})$ when we verify that its Hessian matrix $H\mu_{i}, \mu_{j}$ is positive semi-definite for all feasible $\mu_{i}, \mu_{j}$. This is however not the case since no closed-form solution for $\mu_{i}^{*}, \mu_{j}^{*}$ exists. And to find this optimality, we numerically evaluate our rule-based policy as a solution to the KKT system under varying $\mu_{i}, \mu_{j}$. 

\subsection{Rule-based Queue Policy for Predictive Modeling}
The rule-based policy is basically queue knowledge abstracted as a behavioral model of impatience. From the status information models dispatched and the corresponding tenant reactions evolves a predictive model that encapsulates the probability that a user makes the optimal decision (renege, jockey). Such that, the queue iteratively adjusts it's service rates to minimize the impatience and overall delay in the system. And these performance measures characterize the utility function.
To maximize this utility, the rule-based queue policy determines the next calibration in processing rates at a given interval given input from the predictive model. Essentially, the output from the predictive model is used by the queue to generate action probabilities given the current state in that dispatch interval. At each iteration, the reaction from the tenants is the input to the predictive model for improvement. We compare our rule-based policy to a hedging point that maximizes the expected discounted return as derived by \cite{LinBing}. In this hedging-point technique, the Markov Decision Process is solved numerically by value iteration on a suitably truncated state space. Here, jockeying is defined by monotone switching curves that partition the state space (based on the queue lengths) into regions where the queue serves its own queue or serves a jockeyed job. Essentially, a table of optimal actions (or of the switching curves) is precomputed and the lookup applied at runtime. 

\section{Numerical Results}
\label{sec:results}
In our empirical experiments, we dispatch the information at defined intervals $r \in \{3,5,7, 9\}$ seconds. For each dispatch interval, we simulated over \textit{300 runs} for varying configuration pairs of $\mu_{i}, \mu_{j}$ with different measures of the arrival rates $\lambda: 3\leq\lambda\leq 17$. 

\begin{figure*}[ht] 
 \captionsetup{skip=2pt,belowskip=0pt}
    \centering
            \makebox[\textwidth]{\includegraphics[width=0.825\paperwidth]{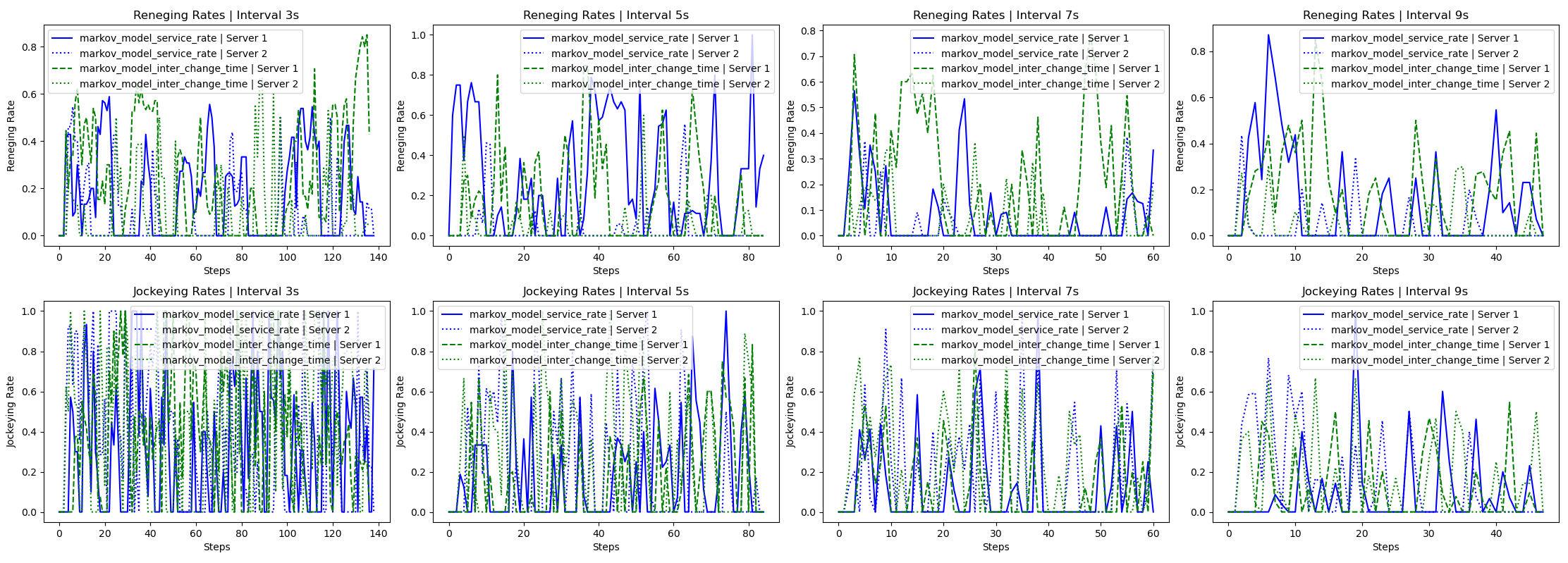}}         
       \caption{\small  When no policy is embedded, the resulting phenomena here is instability regardless of the Markov model of information dispatched. This volatility however as expected reduces when the dispatch intervals are increased. }       
    \label{fig:rene_jock_rates_no_policy_enabled}
\end{figure*}

\begin{figure*}[ht]
  \captionsetup{skip=2pt,belowskip=0pt}
    \centering
   \makebox[\textwidth]{\includegraphics[width=0.825\paperwidth]{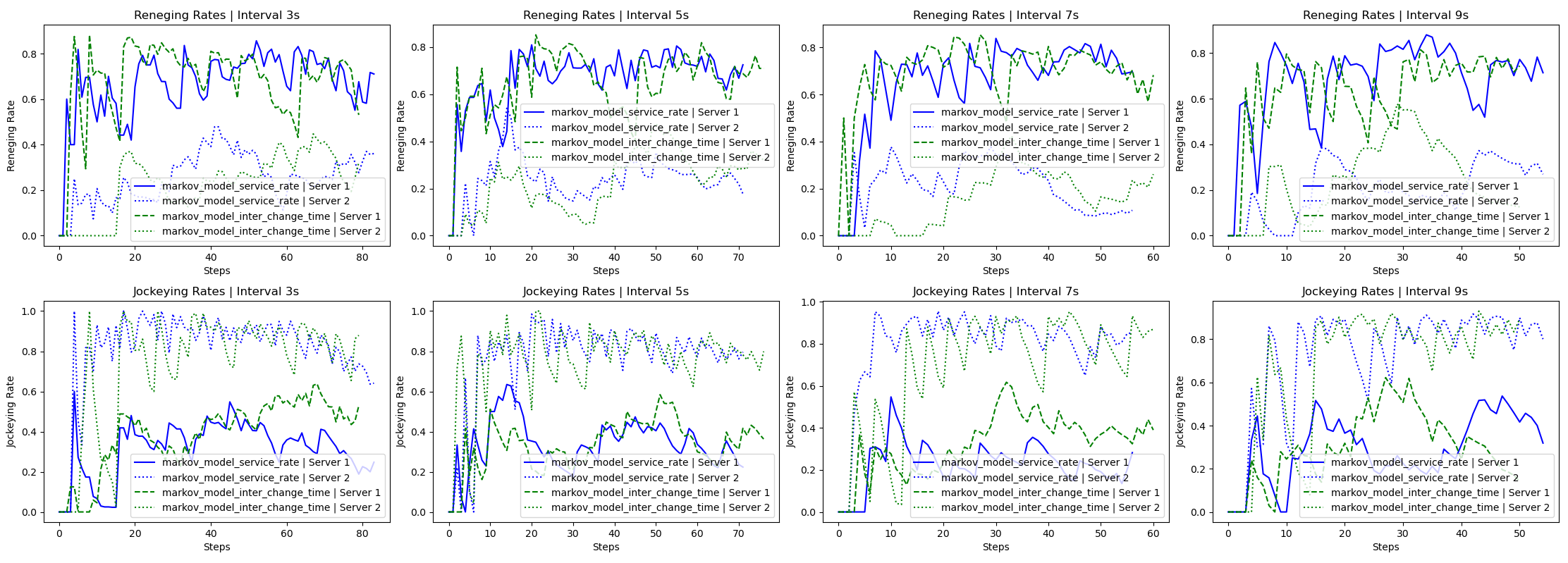}}
      \caption{\small The rule-based queue policy that learns the dispatching and service rates manages to regulate the impatience with the Markov model of the service rates proving more vital especially in keeping the jockeying minimal.}
  \label{fig:rene_jock_rates_policy_enabled}
 \end{figure*}

\begin{figure*}[ht]
\captionsetup{skip=2pt,belowskip=0pt}
      \centering 
      \makebox[\textwidth]{
      \includegraphics[width=0.85\paperwidth]{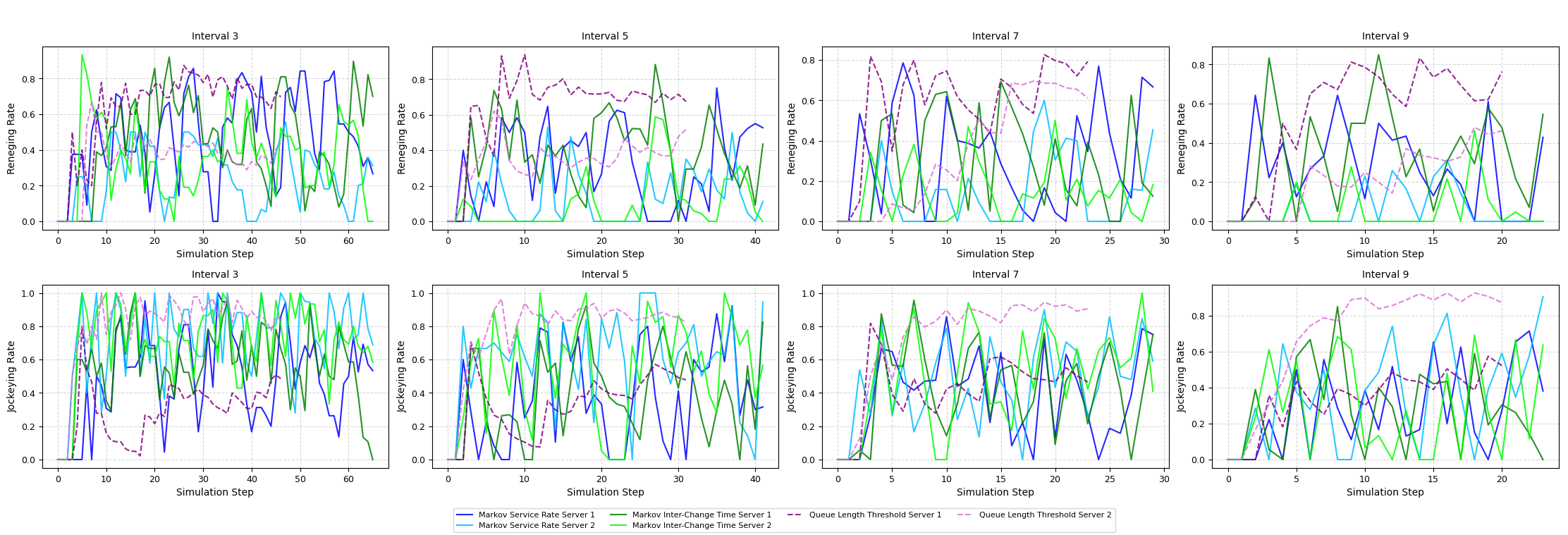}
      }
      \caption{\small Across all intervals, slightly lower and more stable renege rates are observed for both policies, though the hedge-point policy remains relatively higher. The jockeying rates for all policies begin to converge at higher dispatch intervals and show less volatility.}  
      \label{fig:policies}
  \end{figure*}

\noindent Figure \ref{fig:rene_jock_rates_no_policy_enabled} is illustrative of when no server policy is embedded. Here, the chaotic and variability in the profile of the reneging and jockeying rates is evidence of instability. Even under this volatility, it is evident that the Markov model of the service rates yields less impatience in comparison to the Markov model of the changes in the queue length sizes. 
This is logical since with the interchanging times dispatches, requests receive a naive snapshot of the entire dynamics leading to overreacting. This indirectly yields jockeying back and forth to find the short-lived pool. On the other hand, the service rates information provides a direct mapping to how fast jobs are processed in a queue to control the impatience.
Figure \ref{fig:rene_jock_rates_policy_enabled} shows the optimized policy from the predictive modeling where this impatience is more controllable. The frequency of dispatching appears to lead to optimality between the intervals 5 to 7 where the impatience is minimized and stability, particularly when requests get the service rates information. 
Figure \ref{fig:policies} compares the hedging-point policy to the rule-based policy. Although the hedge-point jockeying policy here seems more stable, higher renege and jockeying rates are observed. The  Markov-based policies on the other hand are more volatile, but potentially record lower rates especially in lower intervals.
Increasing the dispatch interval tends to reduce rate volatility and brings all policies closer together, though the hedge-point policy remains smoother.
Ideally, this evaluation reveals that to minimize the volatility and achieve predictable system behavior, the hedge-point policy seems more suitable. But for optimality in the service or lower rates, Markov-based policies may be preferable, especially at lower intervals. 

\begin{figure} 
\captionsetup{skip=2pt,belowskip=0pt}
      \centering      
      \includegraphics[width=0.52\textwidth]{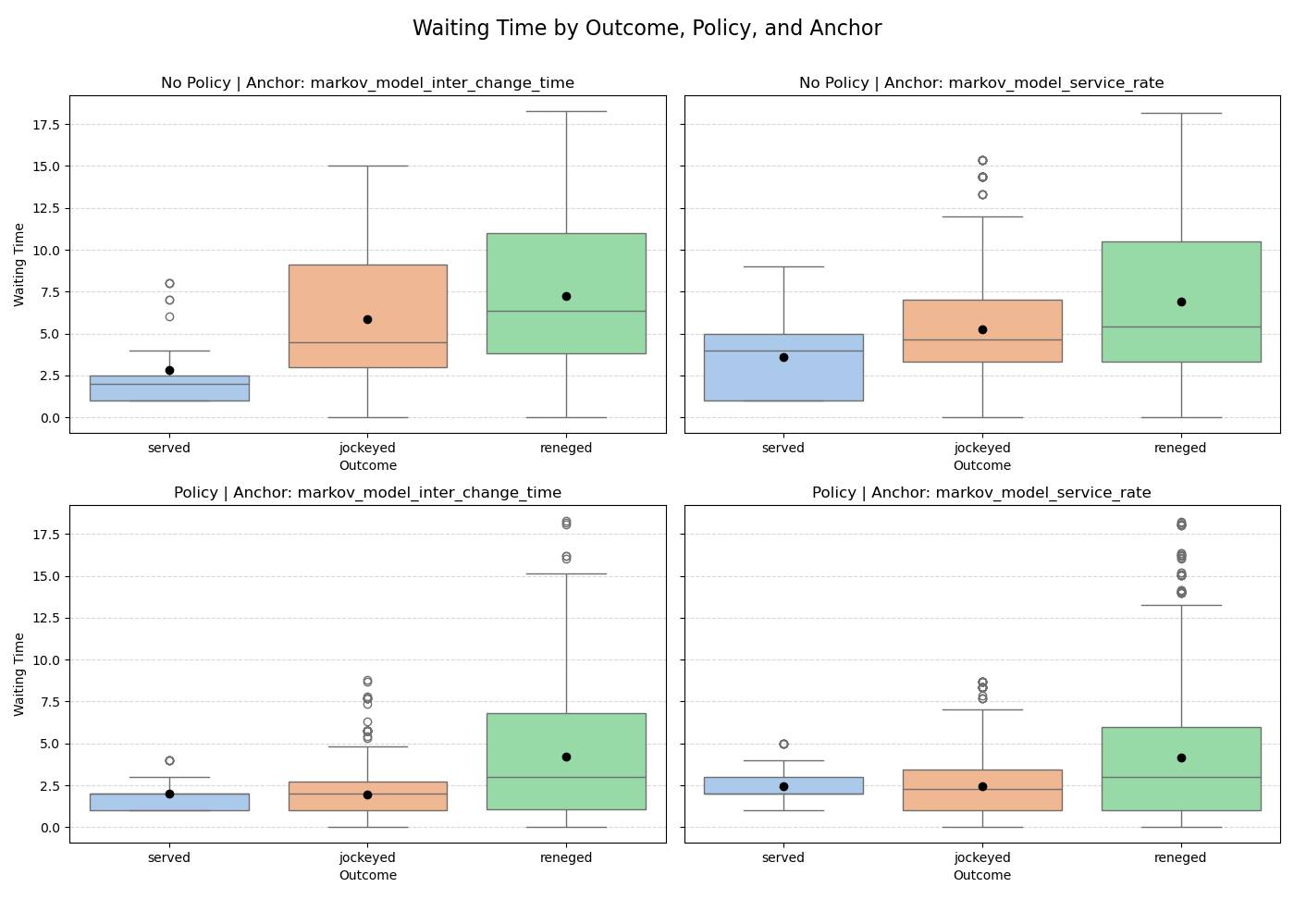} 
      \caption{\small The box plot showing the median waiting time when a policy is embedded versus otherwise for both Markov information models. The waiting time and spread here are high when no policy is embedded for both impatience behavior \protect\cite{kiggundu2025information}.} 
      \label{fig:wait_length}
  \end{figure}

Figure \ref{fig:wait_length} illustrates the overall effect of the information model dispatches on the waiting time of the impatient tenant. When no policy exists in the system, requests that abandon or switch tend to wait for longer as observed from the median measures. The predictive model on the contrary encapsulates useful queue descriptor knowledge which yields optimal waiting times for tenants. The waiting times record lower medians for both reneging and jockeying requests when we embed the policy but more optimality is observable when the Markov model of the service rates is dispatched.

\begin{figure} 
      \centering      
      \includegraphics[width=0.535\textwidth]{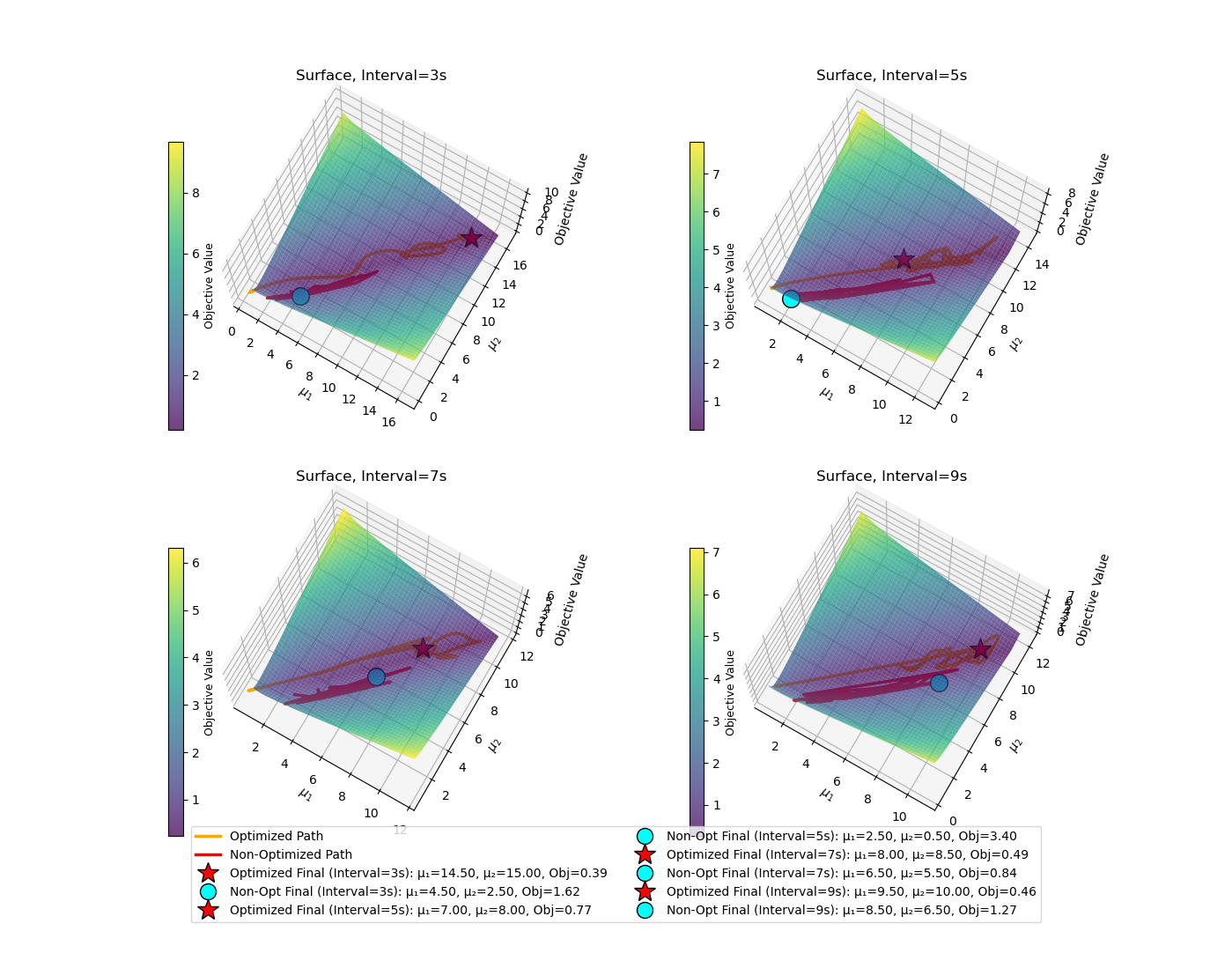}
      \caption{\small Here, the rule-based policy leads to substantially better (lower) objective values in comparison to when no policy is embedded. Our policy finds the near-optimum despite changes in the interval and impatience. Absence of a control policy on the other hand is consistently suboptimal under the behavioral changes since it cannot adapt. } 
      \label{figs:service_optimirieng}
  \end{figure}

Table \ref{tab:results_summary} provides a quantitative summary of Figure \ref{figs:service_optimirieng}. Here, the optimized policy is statistically robust and yields lower yet more consistent objective values (mean = 0.53) in comparison to no optimization at all (mean = 1.78). 
The effect of the intervals is evident from the distance between optimal and non-optimal endpoints as the interval increases (from 3s to 9s). The optimized trajectory in each interval achieves a lower objective value consistently as the landscape changes, finding better solutions, while the non-optimized path sometimes does not. The path is visibly more oriented towards the valleys or optimal regions in the surface unlike in the non optimal case. And the variability is smaller for optimized results (Std. Dev = 0.15) than for non-optimized (Std. Dev. = 0.97), indicating more reliability in performance.  
\begin{table}[ht]
  \centering
  \caption{Optimized vs Non-Optimized Objective Values and Statistical Summary} 
  \rowcolors{2}{LightGray}{white}
  \resizebox{\columnwidth}{!}{%
    \begin{tabular}{c c c c c}
      \toprule
        {\scriptsize \textbf{Interval}} 
      & {\scriptsize \textbf{Opt. Obj.}} 
      & {\scriptsize \textbf{Non-Opt. Obj.}} 
      & {\scriptsize \textbf{Opt.\ $(\mu_{1},\mu_{2})$}} 
      & {\scriptsize \textbf{Non-Opt.\ $(\mu_1,\mu_2)$}} \\
      \midrule
      {\footnotesize 3s}  & {\footnotesize 0.39} & {\footnotesize 1.62} & {\footnotesize (14.50, 15.00)} & {\footnotesize (4.50, 2.50)} \\
      {\footnotesize 5s}  & {\footnotesize 0.77} & {\footnotesize 3.40} & {\footnotesize (7.00, 8.00)}   & {\footnotesize (2.50, 0.50)} \\
      {\footnotesize 7s}  & {\footnotesize 0.749} & {\footnotesize 0.84} & {\footnotesize (8.00, 8.50)}  & {\footnotesize (6.50, 5.50)} \\
      {\footnotesize 9s}  & {\footnotesize 0.46} & {\footnotesize 1.27} & {\footnotesize (9.50, 10.00)} & {\footnotesize (8.50, 6.50)} \\
      \midrule
      \rowcolor{white}
      {\scriptsize \textbf{Mean}}     & {\footnotesize 0.53} & {\footnotesize 1.78} & \multicolumn{2}{c}{{\scriptsize \textbf{Avg.\ Impr.}: 1.26}} \\
      {\scriptsize \textbf{Std Dev}}  & {\footnotesize 0.15} & {\footnotesize 0.97} & \multicolumn{2}{c}{}               \\
      {\scriptsize \textbf{Min}}      & {\footnotesize 0.39} & {\footnotesize 0.84} & \multicolumn{2}{c}{}               \\
      {\scriptsize \textbf{Max}}      & {\footnotesize 0.77} & {\footnotesize 3.40} & \multicolumn{2}{c}{}               \\
      \bottomrule
    \end{tabular}%
  }
  \label{tab:results_summary}
\end{table}

\section{Conclusion and Outlook}
\label{sec:conlusion}
The role of system status information in shaping the decisions of impatient tenants is understudied, yet it is central to realizing data-driven, decentralized resource allocation in next-generation networks. In this work we introduced a queue disclosure approach based on two queue-descriptor information models to guide the rationality of the impatient tenant. We analyzed how the periodic dispatch of such queue state information affects tenant jockeying and reneging. We then developed a lightweight rule-based predictive policy that learns and adapts service rates from tenant responses under potentially stale information. Our numerical study compares a scenario where no policy (baseline) is embedded against the hedging-point MDP benchmark, and the bulletin-driven rule-based policy and quantifies tradeoffs in mean delay, jockeying and reneging rates. The results indicate that both informed policies substantially reduce tenant impatience relative to the no-policy baseline; the hedging-point policy appearing more suitable for stabilizing the system, while the rule-based policy is more robust to information staleness and the nonstationary service dynamics typical of edge cloud environments. Future work would include more information models that abstract the subscription costs and replacing the rule-based heuristics with reinforcement learning techniques. 
Furthermore, the computational and communication overhead of this bulletin mechanism needs to be quantified to achieve clarity about the practical regimes where the bulletin-driven policy is beneficial. A performance evaluation of the policy under bursty, heavy-tailed arrival processes and non-exponential service times is equally relevant to assess the applicability of these concepts beyond the $M/M/2$ Poisson setting. Finally, there is need to scale experiments to multi-queue heterogeneous servers, explore lightweight hierarchical coordination and validate this approach on prototype MEC testbeds with realistic delay traces.

\section*{Acknowledgment}
This work is supported in fully by the German Federal Ministry of Research, Technology and Space \textbf{(BMFTR)} within the \textit{Open6GHub} project under grant numbers \textit{16KISK003K and 16KISK004}.

\begin{acronym}[HRTEM]
  \acro{QoS}{Quality of Service}
  \acro{QoE}{Quality of Experience}
  \acro{AI}{Artificial Intelligence}
  \acro{NSSF}{Network Slice Selection Function}
  \acro{NSSI}{Network Slice Subnet Instance}
  \acro{NSI ID}{Network Slice Instance ID}
  \acro{NWDAF}{Network Data Analytics Function}
  \acro{SMF}{Session Management Function}
  \acro{SDN}{Software Defined Networks}
  \acro{3GPP}{Third Generation Partnership Project}
  \acro{FCFS}{First Come First Server}
  \acro{LCFS}{Last Come First Serve}
  \acro{MDP}{Markov Decision Process}
  \acro{SBA}{Service based Architectures}
  \acro{5G}{Fifth Generation}
  \acro{6G}{Sixth Generation}
  \acro{UE}{User Equipment}
  \acro{UEs}{User Equipments}
  \acro{M/M/C}{Markovian/ Markovian/ number of queues}
  \acro{M/G/C}{Markovian/General/ number of queues}
  \acro{G/G/C}{General/ General/ number of queues}
  \acro{RAN}{Radio Access Network}
  \acro{IoT}{Internet of Things}
  \acro{O-RAN}{Open Radio Access Network}
  \acro{DU}{Distributed Unit}
  \acro{RU}{Radio Unit}
  \acro{CTMC}{Continuous Time Markov Chain}
  \acro{SIRO}{Serve In Random Order}
  \acro{CDF}{Cumulative Distribution Function}
  \acro{S-NSSAIs}{Single – Network Slice Selection Assistance Information}
  \acro{NSSAI}{Network Slice Selection Assistance Information}
  \acro{KKT}{ Karush–Kuhn–Tucker}
  \acro{FSD}{First-Order Stochastic Dominance}
  \acro{ICD}{Inter‐Changing Time Distribution}
  \acro{MEC}{Multi-access Edge Computing}
  \acro{NTN}{Non Terrestrial Networks}
\end{acronym}

\bibliography{references.bib}
\bibliographystyle{IEEEtran}
\end{document}